\documentclass[]{JFM-FLM_Au}
\usepackage{pdfpages}

\usepackage{mathrsfs}
\usepackage[OT1]{fontenc}

\lefttitle{M. Golden}
\righttitle{Journal of Fluid Mechanics}

\title{A catalog of exact steady states in Kolmogorov flow}

\author{Matthew Golden\aff{1,2}}

\affiliation{\aff{1}Theoretical Division, Los Alamos National Laboratory,
Los Alamos, NM 87544, USA
\aff{2}School of Physics, Georgia Institute of Technology, Atlanta, GA 30318, USA}

\corresau{Matthew Golden, \email{mgolden@lanl.gov}}

\begin{document}
\maketitle

\begin{abstract}
Two-dimensional Kolmogorov flow is a common proving ground for the dynamical systems picture of turbulence. Hundreds of unstable periodic orbits have been found for this system across many Reynolds numbers and domain sizes, although the most well explored is $\Rey=40$ on a square $2\pi\times 2\pi$ domain with a forcing wavenumber of $k=4.$ Here I present a catalog of 8952 exact steady states of the flow in this canonical parameter regime. These solutions provide a fixed constellation informing the dynamics of both turbulence and transients.
\end{abstract}

\begin{keywords}
Authors should not enter keywords on the manuscript; these are assigned
during the online submission process.
\end{keywords}

\section{Introduction}
\label{sec:intro}
It has been a community effort over recent decades to realize the dynamical systems picture of turbulence in which the geometry and topology of turbulence is determined by invariant sets: equilibria, periodic orbits, higher dimensional tori, and the stable and unstable manifolds of these solutions. These exact solutions to the Navier-Stokes equations have been called Exact Coherent Structures (ECS) because they are exact solutions to the hydrodynamic model and have historically been used to understand the recurrence of large coherent structures observed in turbulence. This picture can be attributed in part to Eberhard Hopf \citep{hopf1948mathematical} who viewed turbulence as a finite dimensional manifold in the space of admissible flow fields, although we know now that chaotic dynamics tend to produce fractal chaotic sets rather than smooth manifolds \citep{eckmann1985ergodic}.

The primary barrier to the dynamical systems picture is obvious to any practitioner: the numerical computation of ECS is difficult. Any invariant set requires optimization over a high dimensional space requiring significant memory for variational methods \citep{lan2004variational, parker2022variational} or significant walltime for shooting methods \citep{chandler2013invariant}. While steady state solutions are the cheapest to converge, periodic orbits provide the dynamical skeleton of turbulence \citep{cvitanovic1991periodic}. Periodic Orbit Theory (POT) predicts that ergodic measures of strange attractors can be expressed as weighted sums of unstable periodic orbits that densely cover the attractor \citep{Cvitanovic2020Chaos}. 

Kolmogorov flow was suggested by Andrey Kolmogorov in 1959 \citep{arnold1959seminar}. A fluid with periodic boundary conditions is forced monochromatically to drive various cascades. 2D Kolmogorov flow has been of wide interest to dynamical systems researchers given its dynamical complexity and simple implementation \citep{
platt1991investigation,
inubushi2012covariant,
chandler2013invariant, 
suri2014velocity,
lucas2014spatiotemporal,
lucas2015recurrent,
farazmand2016adjoint,
farazmand2016dynamical,
tithof2017bifurcations,
suri2017forecasting,
farazmand2017variational,
suri2018unstable, 
farazmand2019closed,
suri2019heteroclinic,
parker2022variational,
lucas2022stabilization,
de2023data,
constante2024data,
vela2024large,
page2025computation,
engel2025search,
cleary2025dynamical,
vinograd2026dimensional,
parker2026synthetic,
zhigunov2026exact,
beck2026identifying}.
\footnote{Any attempt to list dynamical systems research on Kolmogorov flow will surely be incomplete.}

A common theme of these investigations is the convergence of periodic orbits; these solutions shadow turbulence and give insight into the underpinning nonlinear processes of the flow. At this time, hundreds of periodic orbits are known by several active research groups. These are usually identified by seeding a Newton-Raphson algorithm with near recurrences from turbulence. While these loops provide a \textit{dynamical skeleton} of the chaotic set, the simpler set of steady states provides a \textit{constellation}. These steady states are fixed points, and they constrain the topology and geometry of the flow. Three such solutions were provided in \cite{chandler2013invariant} and twenty-five were later provided in \cite{farazmand2016adjoint}. Here I provide a constellation orders of magnitude larger: 8952 steady states identified in 2D Kolmogorov flow at $\Rey=40$.

\section{Numerical methods}
Kolmogorov flow is a choice of geometry and forcing for the incompressible Navier-Stokes equation
\begin{align}
    & \nabla \cdot {\bf u} = 0,\\
    &\partial_t {\bf u} + {\bf u} \cdot \nabla {\bf u } + \nabla p = \frac{1}{\Rey} \nabla^2 {\bf u} + {\bf f}.
    \label{eq:NS}
\end{align}
I will work with $\Rey=40$, ${\bf x} \in [0, 2\pi)^2$ with periodic boundary conditions, and ${\bf f} = \sin(4y) \hat{i}$ following \cite{chandler2013invariant}. It is advantageous to work in the vorticity-velocity formulation defining $\omega \equiv \partial_x u_y - \partial_y u_x$ to obtain the pressure-free vorticity equation
\begin{equation}
    \partial_t \omega + {\bf u} \cdot \nabla \omega = \frac{1}{\Rey} \nabla^2 \omega - 4 \cos(4y).
\end{equation}
The vorticity field is stored numerically, and the velocity field is obtained by solving the simple linear system $\partial_x u_x + \partial_y u_y = 0$ and $\partial_x u_y - \partial_y u_x = \omega$ in Fourier space. The mean flow is undetermined by these constraints, so I enforce $\langle {\bf u} \rangle  = 0$ at all times, where 
\begin{equation}
    \langle \mathcal{O} \rangle = \frac{1}{(2\pi)^2} \int_0^{2\pi} dx  \int_0^{2\pi} dy \, \, \mathcal{O}.
\end{equation}
The vorticity forcing $\cos (4y)$ has continuous translation symmetry in $x$, so this system supports steady state traveling waves satisfying $(\partial_t + c \partial_x) \omega = 0$. Combining this wave equation and the vorticity equation gives the spatial constraint
\begin{equation}
F(\omega, c) = -c\partial_x \omega + {\bf u} \cdot \nabla \omega - \frac{1}{\Rey} \nabla^2 \omega + 4 \cos(4y) = 0.
\label{eq:TW}
\end{equation}
Note that equilibria $\partial_t \omega = 0$ are just a special case of traveling wave with $c=0$. This equation is the primary concern of this manuscript. I discretize this equation pseudospectrally on a $128\times128$ grid with Fast Fourier Transforms (FFTs). The derivatives and velocity evaluation are done in Fourier space with the quadratic nonlinearity being computed by pointwise multiplication followed by 2/3rds dealiasing \citep{orszag1971elimination} to preserve exact continuous translational symmetry. 

An initial guess for a pair $(\omega, c)$ solving Equation \eqref{eq:TW} is converged with Newton-GMRES. A phase condition is added to the Jacobian to prevent motion along the trivial marginal direction $\partial_x \omega$. A preconditioner of $\Rey \nabla^{-2}$ was found critical to the success of GMRES \citep{saad1986gmres}. The Newton step $(\delta \omega, \delta c)$ is obtained by solving 
\begin{equation}
    \begin{pmatrix}
        \frac{1}{\nu \nabla^2}\partial_\omega F & \frac{1}{\nu \nabla^2}\partial_c F \\
        \partial_x \omega & 0
    \end{pmatrix}
    \begin{pmatrix}
        \delta \omega \\ \delta c
    \end{pmatrix} = 
    \begin{pmatrix}
        \frac{1}{\nu \nabla^2}F \\ 0
    \end{pmatrix}.
\end{equation}
This search deviates from the literature in two important ways.
\begin{enumerate}
    \item \textbf{Initial guesses for traveling waves were never seeded from turbulence}. Instead, initial guesses were Fourier filtered random noise with zero mean rescaled to have a standard deviation of $\sigma_\omega \sim 1$; the wave speed $c$ was drawn from a normal distribution with zero mean and $\sigma_c \sim 0.1$. The cutoff in Fourier space was varied from a maximum wavenumber of 4 to 6. All initial guess parameters were dynamically updated by Claude to maximize the probability of success.

    \item \textbf{The Newton-GMRES procedure was designed to be lazy.} Generating a new initial guess is effectively free, so I encouraged abandoning Newton iteration if solutions did not show signs of rapid convergence. This is akin to buying a new lottery ticket as soon as your current ticket seems unlikely to win. This attitude was inspired by conversations at the long program \textit{Multi-Fidelity Methods for Fusion Energy} hosted by the Institute for Pure and Applied Mathematics (IPAM). The initial guesses are converged in single precision (float32) with an adaptive GMRES tolerance of $\sqrt{\| F\| / \|F_0\|}$ clamped between $0.01$ and $0.5$. If the solution did not reach $\| F\| < 10^{-3}$ in 32 Newton-GMRES iterations, then the initial condition is declared a failure and discarded. If single precision iteration does reach this $10^{-3}$ target, then a double precision (float64) Newton-GMRES is kicked off with a fixed GMRES tolerance of $10^{-6}$. If the state reaches a maximum pointwise error of less than $ 5 \times 10^{-13}$ in 80 Newton steps, it is accepted as a candidate solution.
\end{enumerate}

This Newton-GMRES pipeline for Kolmogorov flow was entirely written with Claude Code in C, but the discovered solutions were carefully hand-validated in MATLAB and Python. While Claude rapidly produced source code, there was a nontrivial debugging stage: I had to confront Claude over inconsistent $(x,y)$ ordering, incorrect dealiasing, and failure to set the mean vorticity to zero as required by periodic boundary conditions. While Claude made this project possible and more enjoyable, it is my opinion that current LLMs are not capable of fully autonomous research.

The resulting C code was ran in parallel on 8 cores for three months on an ASUS Expertbook with an Intel Core Ultra 5. This pipeline was migrated to an Ampere Altra Q64 and ran on all 64 cores for an additional month. No GPU was ever used to converge new solutions, although an NVIDIA V100 computed the linear stability of all solutions. I am choosing to halt the search for steady states now as it is increasingly impractical to converge new solutions.

\begin{figure}[h!]
    \centering
    \includegraphics[width=\textwidth]{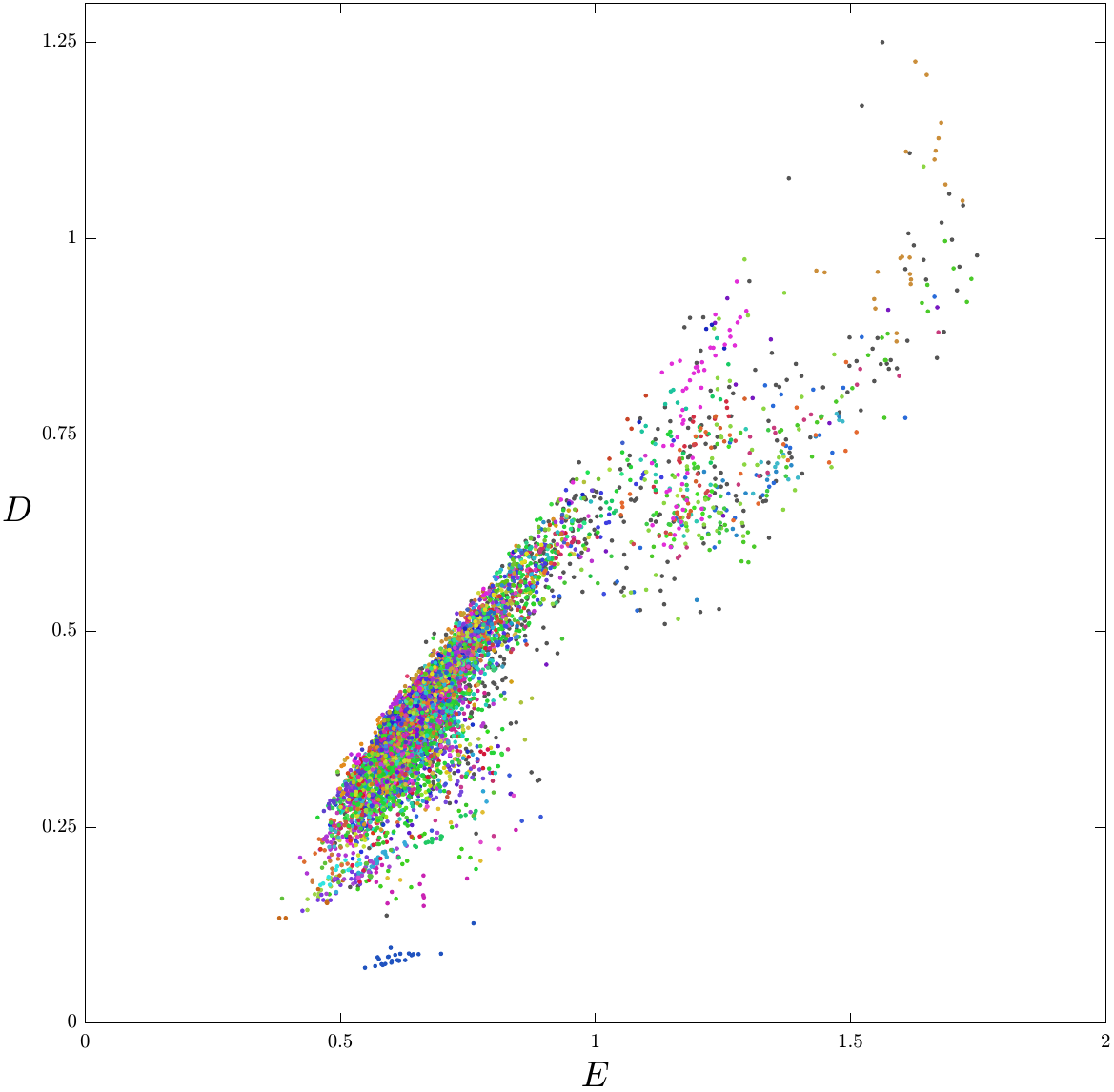}\\
    \caption{Low dimensional projection of all steady state solutions. All 416 families are shown in different colors, with the noise solutions in gray. Family 1 can be seen as the low dissipation cluster in blue.}
    \label{fig:2D_proj}
\end{figure}

\section{Results}
\label{sec:results}

The final catalog contains a total of 8952 steady state solutions comprised of 847 equilibria $(c=0)$ and 8105 traveling waves $(c\neq 0)$. Solutions are considered distinct if after optimizing over symmetries, the maximum pointwise difference in vorticity is greater than $10^{-2}$. There are many solutions near bifurcations, so it is likely that there are true solutions that fall below this discrimination threshold and were ignored as duplicates.

These solutions were clustered with recursive application of the HDBSCAN algorithm \citep{campello2013density} implemented in scikit-learn. The metric used for clustering is the symmetry optimized $L^2$ distance between solutions. Both the 16 discrete symmetries \citep{chandler2013invariant} and continuous translation are optimized over. This distance is a true metric so the optimized distance satisfies the triangle inequality. I use cluster\_selection\_method="leaf" rather than the scikit-learn default "eom", which favors merging into a small
number of large clusters. Because a single HDBSCAN pass leaves many solutions without a cluster assignment, I
recursively re-cluster whatever remains unassigned at each level until a pass finds no further structure, which
occurred after 39 levels. This procedure yields 416 families, containing 8294 of the 8952 solutions (92.6\%); the
  remaining 658 (7.3\%) are not assigned to any cluster. I do not force these into the nearest family: HDBSCAN is a
  density-based method for which leaving outliers unclassified is expected behavior, not a shortcoming.  The single largest family (628 members) does not itself subdivide further under the same procedure at any tested
  threshold. 
  
  This clustering is not intended to be definitive. Rather, it is a guide for interactive exploration of these solutions. Solutions are named according to their family. For example, the seventh member of family 101 is named F101-7. The solutions which could not be clustered instead have the prefix N for ``noise''. Families are ordered by the mean number of unstable directions. Family 1 has the smallest number of unstable directions on average. The ordering within a family is the order in which solutions were discovered by the search. 

\begin{figure}[h!]
    \centering
    \includegraphics[width=\textwidth]{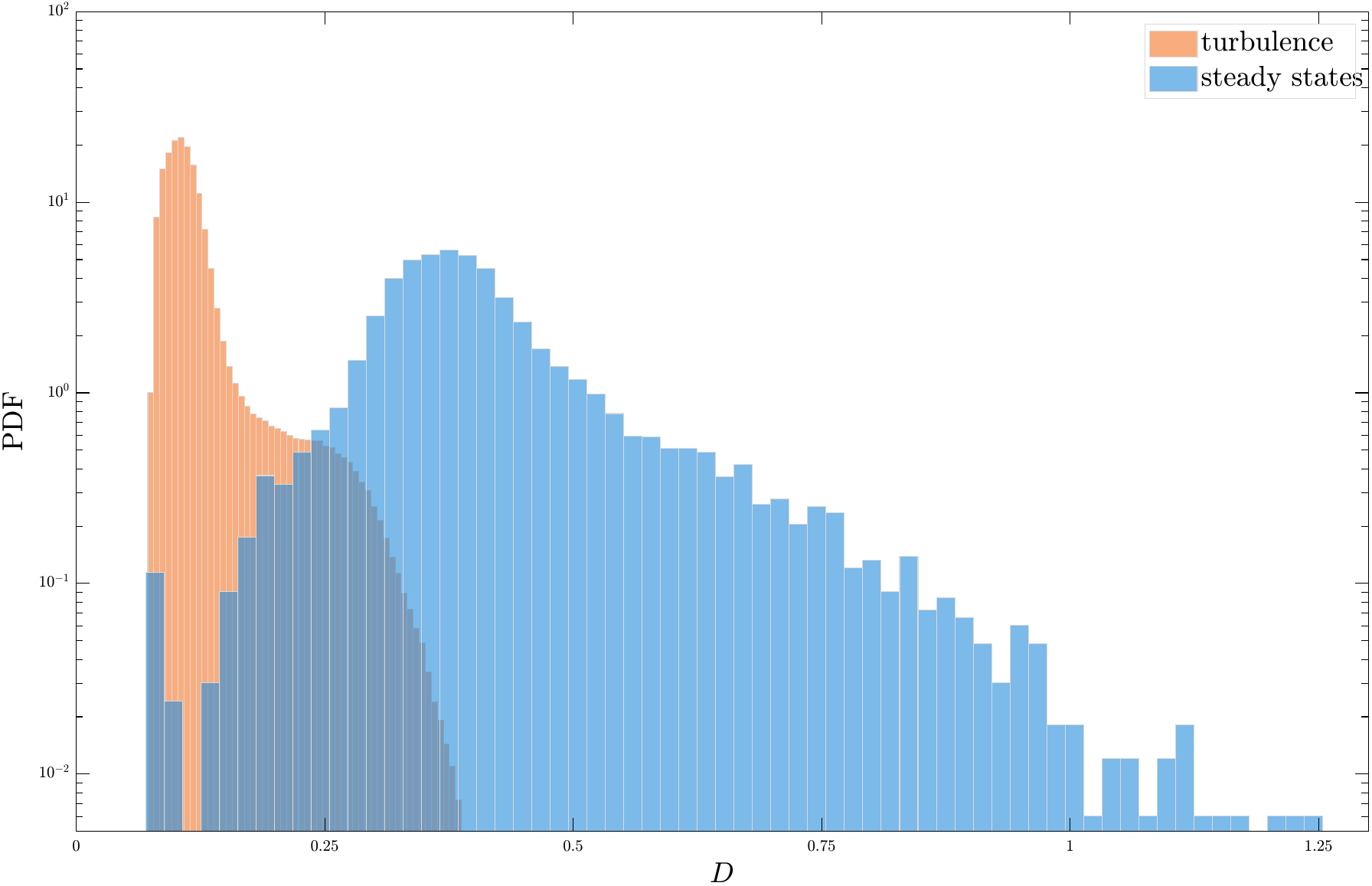}\\
    \caption{A histogram of dissipation for the steady state solutions in the catalog and a turbulent trajectory. Note that the catalog extends to dissipations that are larger than seen in the chaotic set.}
    \label{fig:diss_hist}
\end{figure}

  Figure \ref{fig:2D_proj} shows a 2D projection of all solutions. This projection uses the physical quantities of kinetic energy $E$ and dissipation $D$
  \begin{align}
      & E = \frac12 \langle {\bf u} \cdot {\bf u} \rangle, \\
      & D = \frac{1}{\Rey}\langle \nabla {\bf u} : \nabla {\bf u} \rangle = \frac{1}{\Rey} \langle \omega^2 \rangle. 
  \end{align}
  Figure \ref{fig:2D_proj} shows that the majority of families cannot be meaningfully distinguished by $(E,D)$ alone. An outlier is Family 1, which is shown as a low dissipation cluster in blue. This family is relevant to turbulence, as seen in Figure \ref{fig:diss_hist}.

Figure \ref{fig:diss_hist} shows a histogram of dissipation $D$ for both the catalog of steady states and a turbulent trajectory. Clearly the catalog extends to dissipation values far larger than typically seen in the chaotic set. Such solutions are relevant for transients, but not late time turbulence. Further, there is a clear gap in $D$ for steady state solutions. This gap is physical and is where turbulence spends most of its time. This region of state space is filled with periodic orbits.

\begin{figure}[h!]
    \centering
    \includegraphics[width=\textwidth]{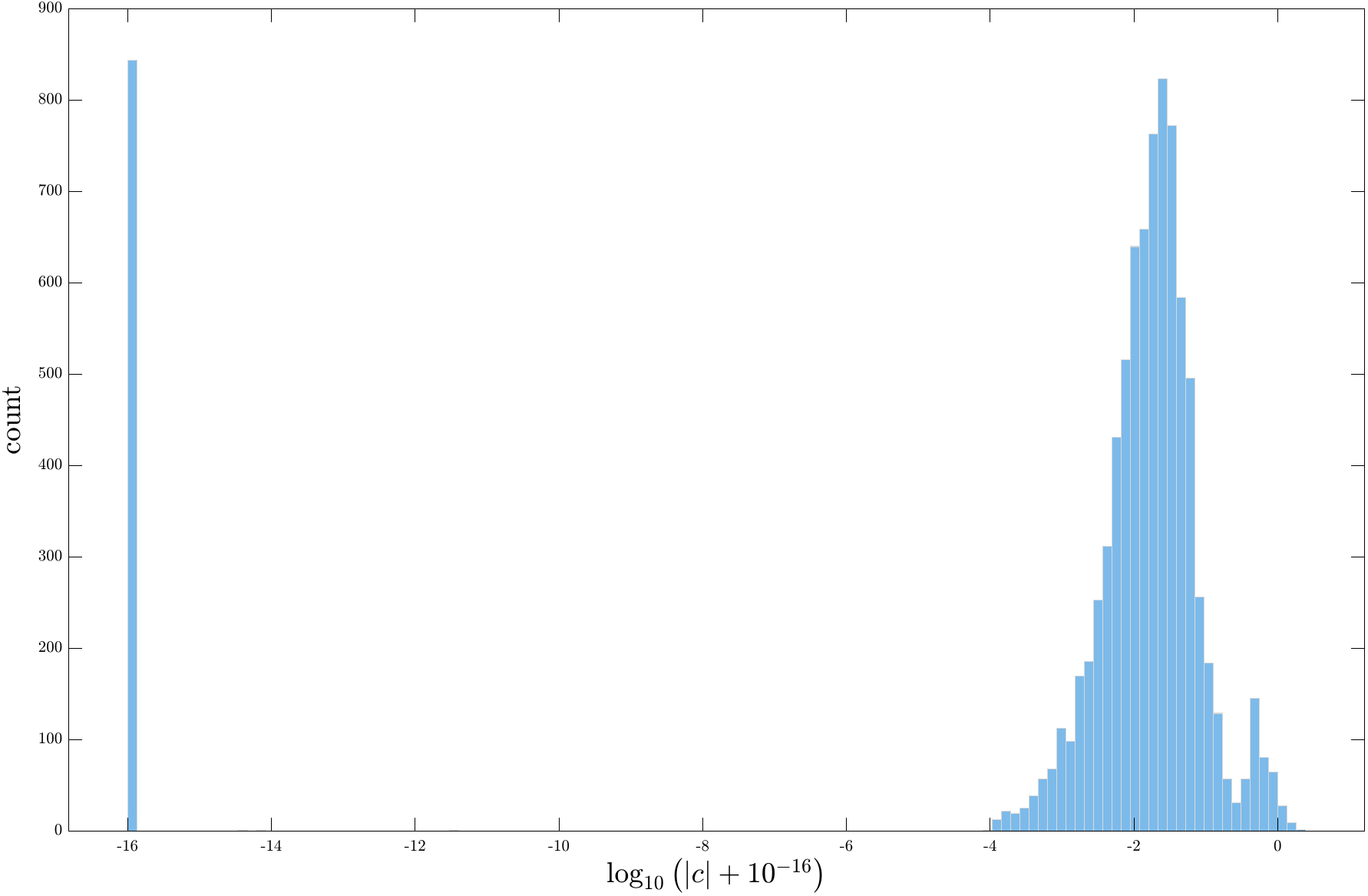}\\
    \caption{A histogram of wave speeds. A factor of $10^{-16}$ is added so that equilibria $(c=0)$ can appear.}
    \label{fig:c_hist}
\end{figure}

\begin{figure}[h!]
    \centering
    \includegraphics[width=\textwidth]{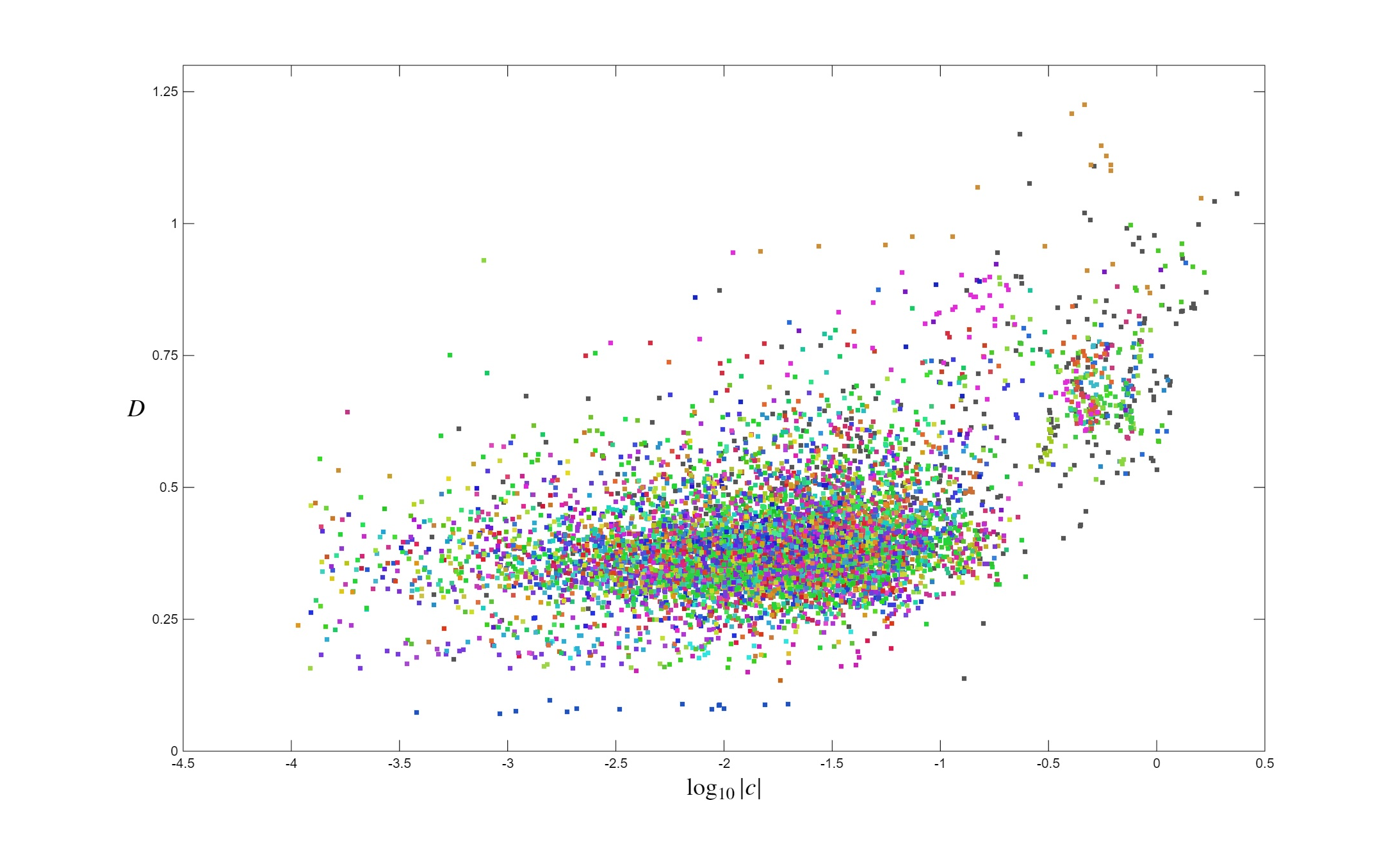}\\
    \caption{Low dimensional projection of dissipation $D$ and wave speed $|c|$. The cluster of high wave speed solutions is associated with higher dissipation. Color corresponds to family. Again family 1 is visible in blue as low dissipation outliers.}
    \label{fig:c_diss}
\end{figure}

Wave speeds were found over the range $0 \leq |c| \leq 2.3511$. A histogram of these speeds is shown in Figure \ref{fig:c_hist}. Note that there is a clear separation between traveling waves and equilibria. The distribution of $c$ does not smoothly approach $c=0$. Further, the distribution of $c$ appears to be bimodal. A small population of high speed solutions is visible. Figure \ref{fig:c_diss} shows how these wave speeds correlate with dissipation. Low $|c|$ solutions have no significant dependence on $D$; however, the high $|c|$ solutions are associated with high $D$. 

\subsection{Linear stability}
\label{sec:stability}

\begin{figure}[h!]
    \centering
    \includegraphics[width=\textwidth]{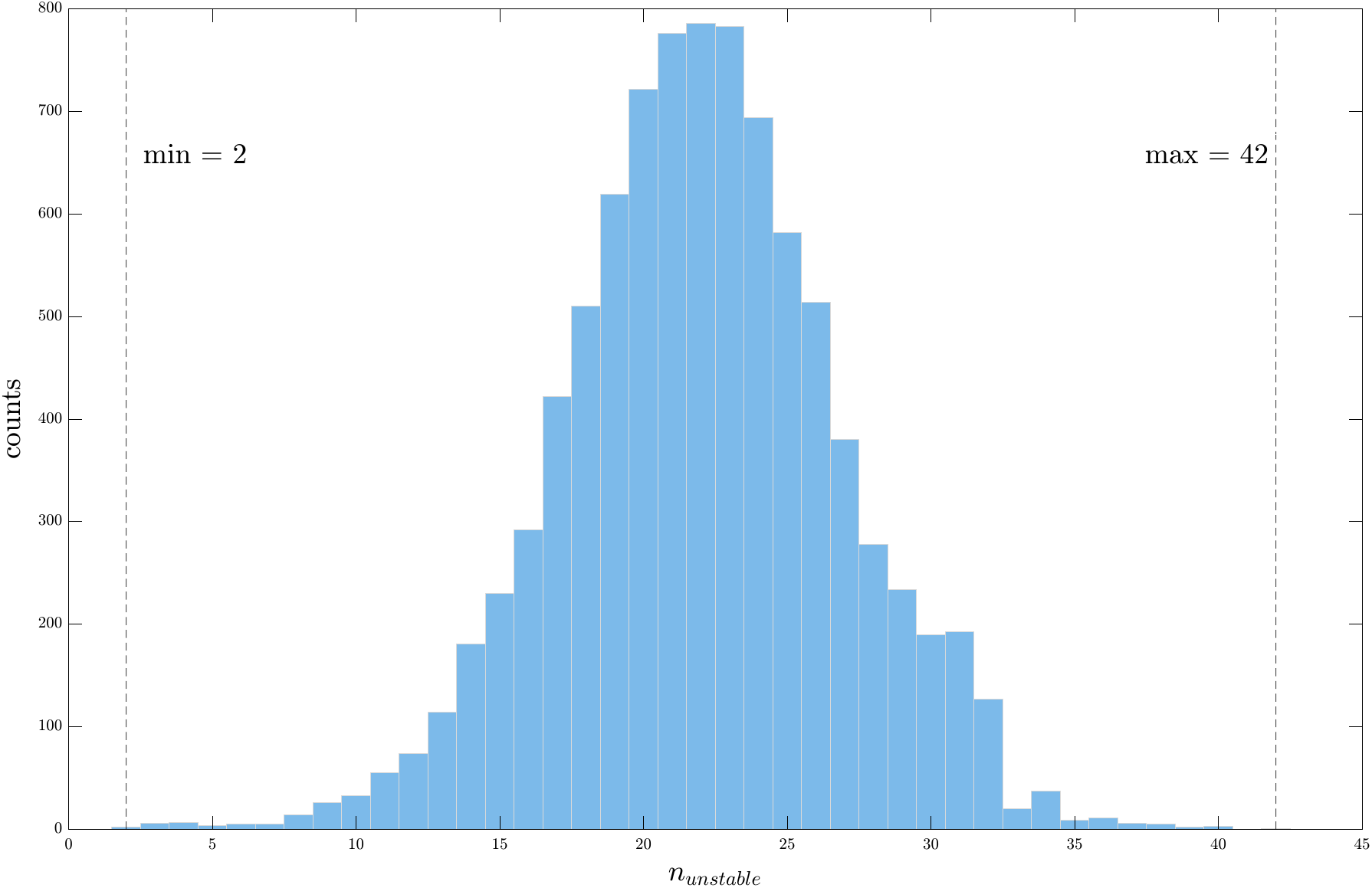}\\
    \caption{A histogram of unstable directions for all solutions in the catalog. No solution was found to be linearly stable.}
    \label{fig:n_unstable}
\end{figure}

For every one of the 8952 solutions I computed the leading 32 eigenvalues and real Schur vectors of the operator
obtained by integrating the linearized vorticity equation to $T=1$. This integration was performed with Lawson-RK4 \citep{lawson1967generalized} in the
solution's comoving frame. If the 32nd vector corresponded to an unresolved complex conjugate pair, then it was discarded. If the number of unstable directions was larger than 32, a larger set of 64 eigenvectors was converged. All solutions had fewer than 64 unstable modes. The real Schur decomposition of this operator was estimated with a restarted block Arnoldi--Krylov--Schur method (block size 40, 8 blocks per cycle, up to 10 restart cycles, Ritz-pair residual tolerance $10^{-13}$). Figure \ref{fig:n_unstable} shows a histogram of unstable directions with a minimum of 2 and maximum of 42 with a mean of 22.0087.

\begin{figure}[h!]
    \centering
    \includegraphics[width=0.9\textwidth]{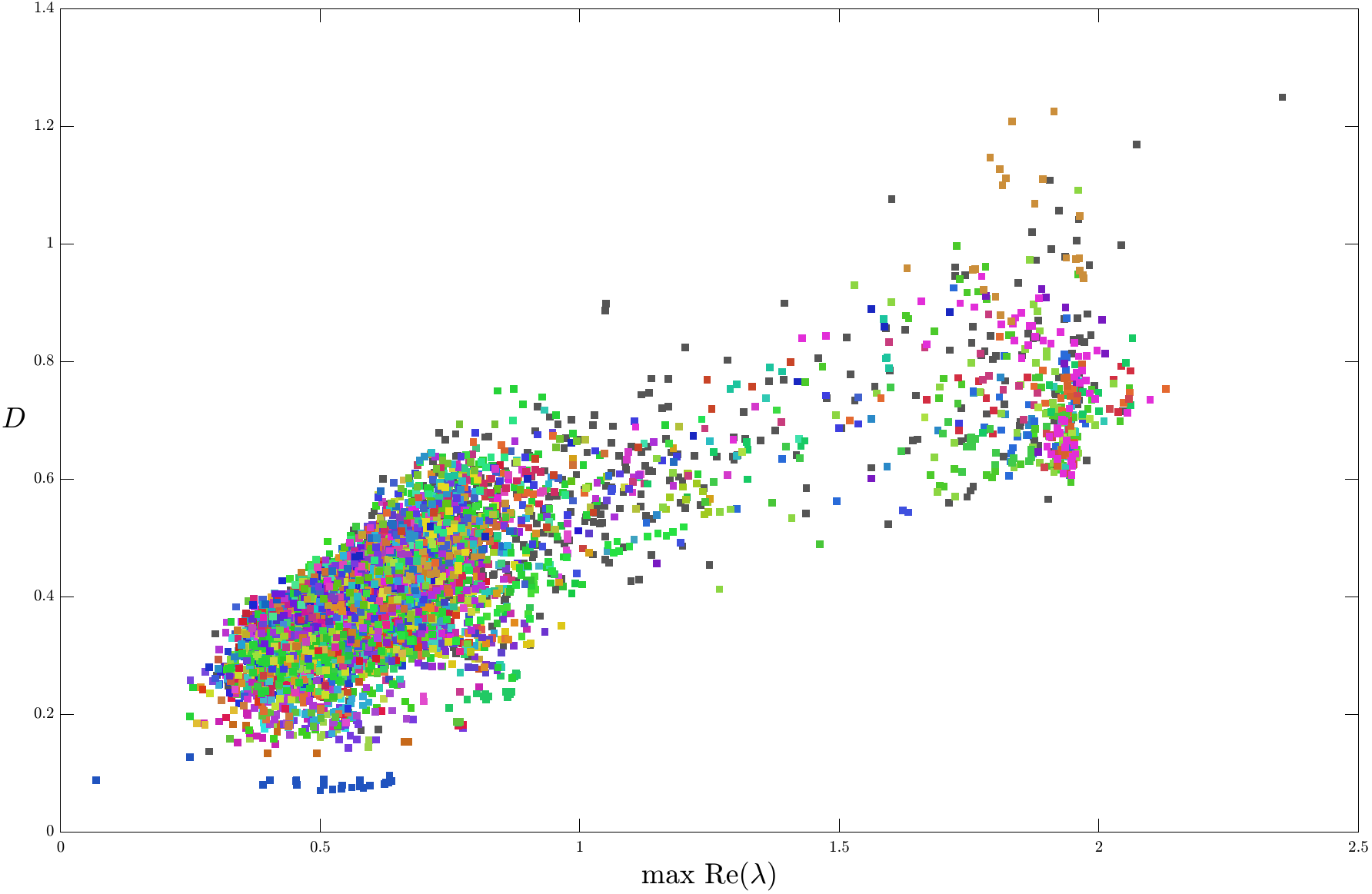}\\
    \caption{Leading linear growth rate versus dissipation. The cluster of highly unstable solutions is associated with higher dissipation. Again Family 1 is visible as a low dissipation outlier in blue. }
    \label{fig:growth_v_dissipation}
\end{figure}

Finite time Floquet multipliers $\mu$ were computed, so only the real part of the continuous time eigenvalues $\partial_t \delta \omega = \lambda \delta \omega$ can be estimated reliably with $\textrm{Re} (\lambda) = T^{-1}\ln |\mu|$. The maximum growth rate $\textrm{max Re}(\lambda)$ is likely dominant under a random perturbation. This growth rate is shown against dissipation in Figure \ref{fig:growth_v_dissipation}. There is a cluster of highly unstable solutions corresponding to higher dissipation.

\subsection{Dynamical relevance to turbulence}
A turbulent trajectory was simulated for $2^{16}$ timesteps with $\Delta t = 0.1$. Each point was compared to the catalog to find the nearest steady state after optimizing over all 16 discrete symmetries and continuous translation. To accomplish this, the first solution in each family was used to create a pivot table: a $416\times 416$ matrix of distances between pivots and a $416 \times 8952$ matrix of distances between pivots and the entire catalog. Since the symmetry optimized distance satisfies the triangle inequality, I can quickly determine the closest steady state to a turbulent snapshot by only optimizing distance for states that the triangle inequality does not eliminate.

\begin{figure}[th!]
    \centering
    \includegraphics[width=0.49\textwidth]{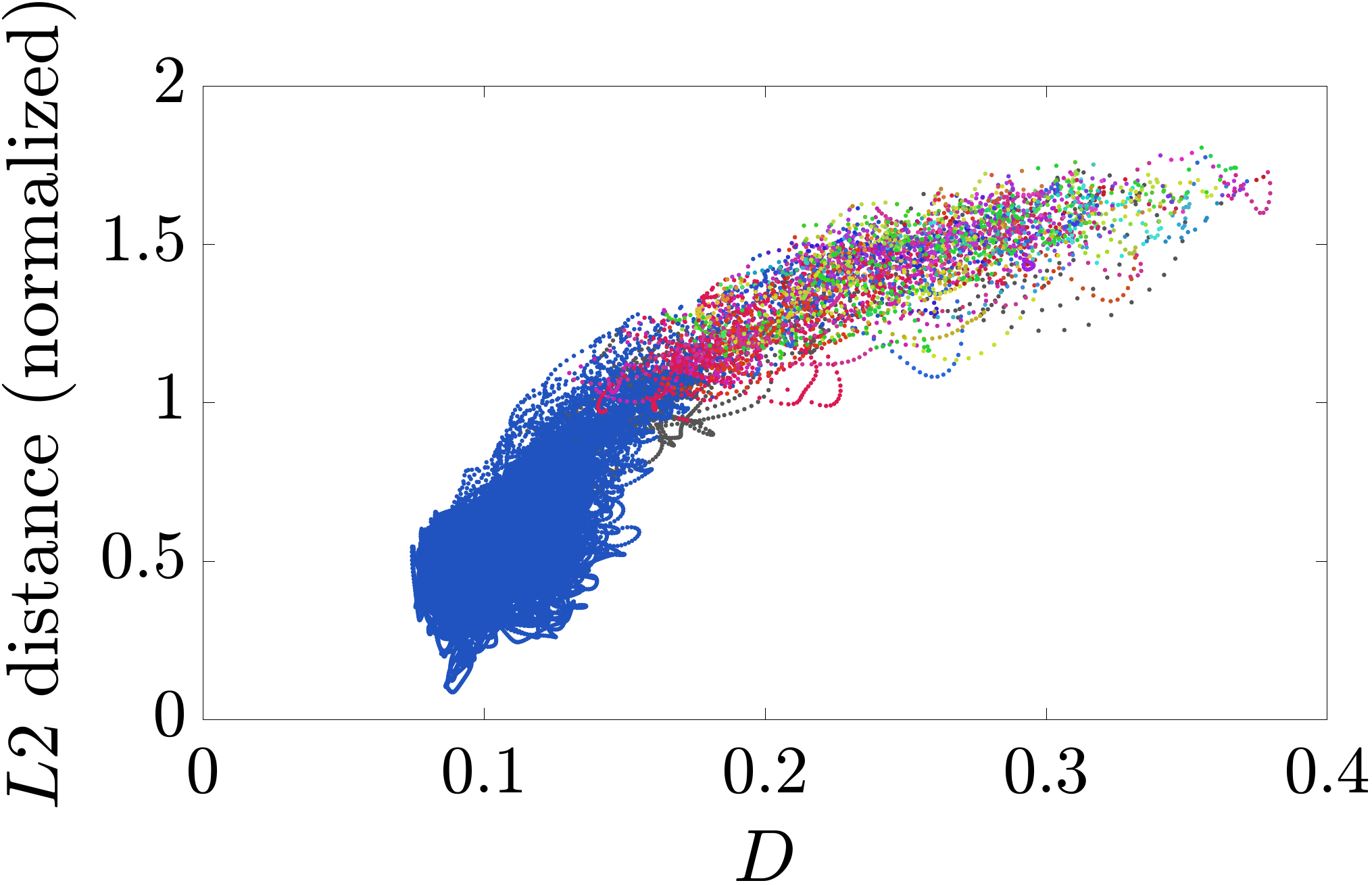}
    \includegraphics[width=0.49\textwidth]{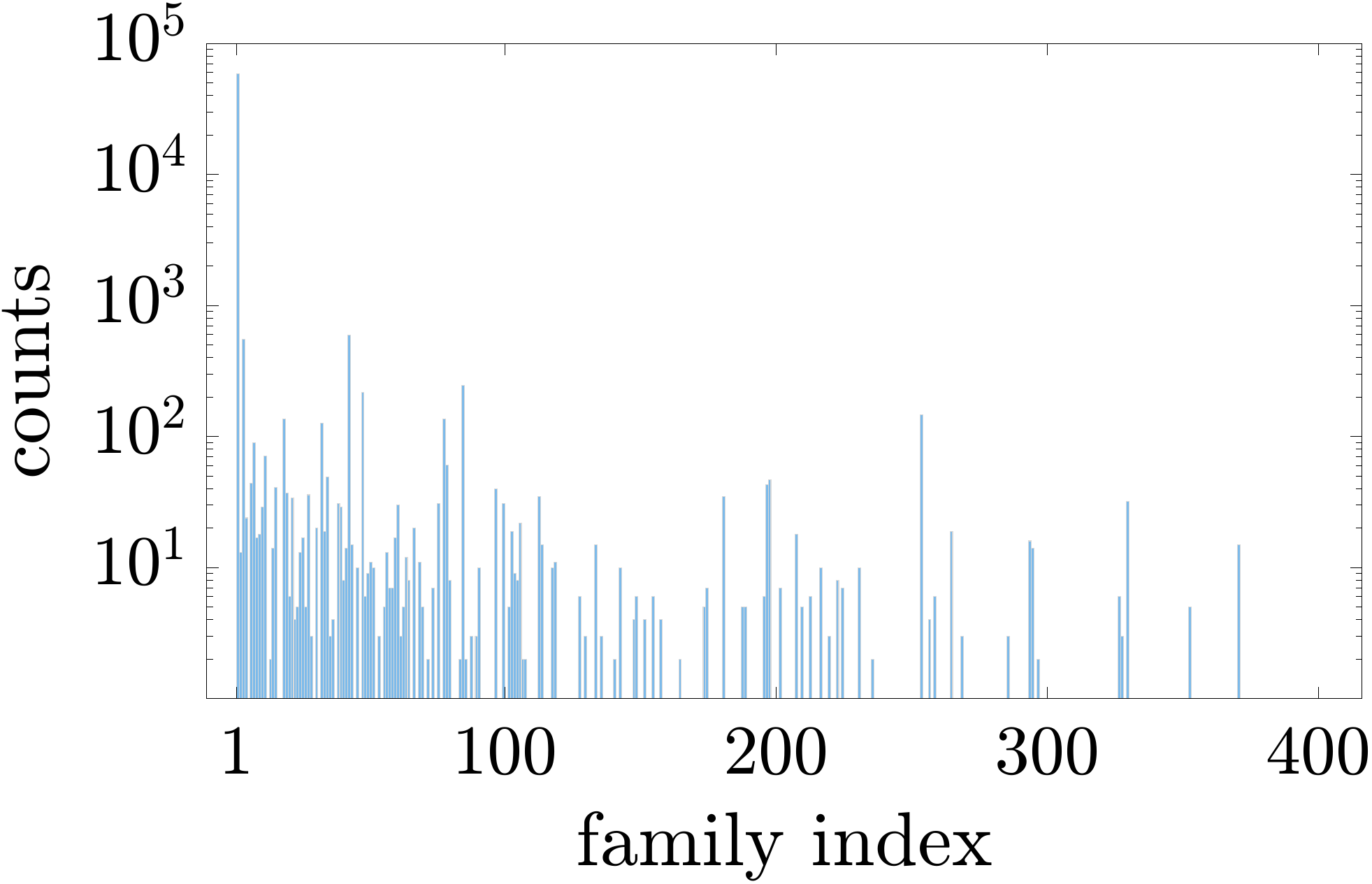}\\
    \caption{(a) $L^2$ distance to the closest steady state for a turbulent trajectory. This distance is normalized by the $L^2$ norm of solution F1-1. Each color corresponds to a different family, with Family 1 being the large blue core at low dissipation. (b) Histogram of nearest family to a turbulent state. Note that Family 1 is most relevant by an order of magnitude. More unstable families are visited less frequently, with Families 371-416 not visited at all by this turbulent segment. More than 90\% of turbulent states are closest to Family 1.}
    \label{fig:distance}
\end{figure}

Figure \ref{fig:distance}(a) shows the normalized distance from turbulence to the nearest steady state in the catalog. The color corresponds to the family of the nearest steady state. At low dissipation, Family 1 dominates. Other families only become relevant when $D \gtrsim 0.2$.  Figure \ref{fig:distance}(b) shows the frequency of visiting each family. Family 1 dominates with 90.26\% of turbulent states being closest to this low dissipation cluster. This is unsurprising, as Family 1 is both low in unstable directions and in $\textrm{max Re}(\lambda)$ (see Figure \ref{fig:growth_v_dissipation}). 

Figure \ref{fig:distance}(b) may give the impression that a sparse subset of families are closest to turbulence. This used a turbulent segment of length $t \approx 6{,}500$. The rarity of extreme high dissipation events makes it difficult to quantify which families are truly relevant to extreme behavior. A much longer turbulent segment would be required to make such statements. Rather, Figure \ref{fig:distance}(b) suggests that a wide range of families are relevant to extreme events. However, the $L^2$ distance to steady states during extreme events can be $\sim 3$ times larger than the distance during the shadowing of Family 1.

\section{Discussion and conclusion}
\label{sec:discussion}

Running an LLM-driven Newton--GMRES search produced a vast catalog of 8952 steady state solutions of 2D Kolmogorov flow. These solutions have been clustered into 416 families using a symmetry optimized $L^2$ distance. No forward time integration was used to converge these solutions, so they span both the dissipation regimes of turbulence and bizarre high dissipation regimes only seen in initial transients. These steady states form a fixed constellation rather than a dynamical skeleton of turbulence, but they are valuable landmarks for anyone interested in the dynamical systems picture of Kolmogorov flow.

This work would not have been possible without Claude Code. The entire pipeline was written in C by agents and carefully debugged and validated by me. The agents had several bugs and struggled with maintaining conventions across the codebase. They were not capable of fully autonomous research, but could achieve high quality solutions with guidance.

\begin{bmhead}[Data availability statement.]
The full solution database \texttt{solutions.h5} (1.09GB) is available at
\url{https://huggingface.co/datasets/mgolden30/Kolmogorov_TW}. This dataset along with the linear stability ($\sim$38GB) is available at \url{https://invariant-sets.org}.
\end{bmhead}

\begin{bmhead}[Acknowledgements.]
I am grateful to the UCLA Institute for Pure and Applied Mathematics (IPAM), which allowed me to discuss this work with George Holt, Phil Travis, Clarence Rowley, and many others. 
I am grateful for helpful discussions with Predrag Cvitanovi\'c, Roman Grigoriev, Mike Schatz, and Dmitriy Zhigunov.
\end{bmhead}

\begin{bmhead}[Funding.]
Part of this research was performed while the author was visiting the Institute for Pure and Applied Mathematics (IPAM), which is supported by the National Science Foundation (Grant No. DMS-1925919/2422832).
\end{bmhead}

\begin{bmhead}[Declaration of interests.]
The author reports no conflict of interest.
\end{bmhead}

\bibliographystyle{jfm}
\bibliography{jfm}

\newpage
\begin{appen}
\section{Representative samples}\label{appA}
Visualizing all 8952 solutions in a PDF document is unwise. Here I provide a random set of 400 solutions in Figure \ref{fig:rand400} and a complete visualization of Family 1 in Figure \ref{fig:Fam1}. All solutions can be interacted with at \url{https://invariant-sets.org}.

\begin{figure}[th!]
    \centering
    \includegraphics[width=\textwidth]{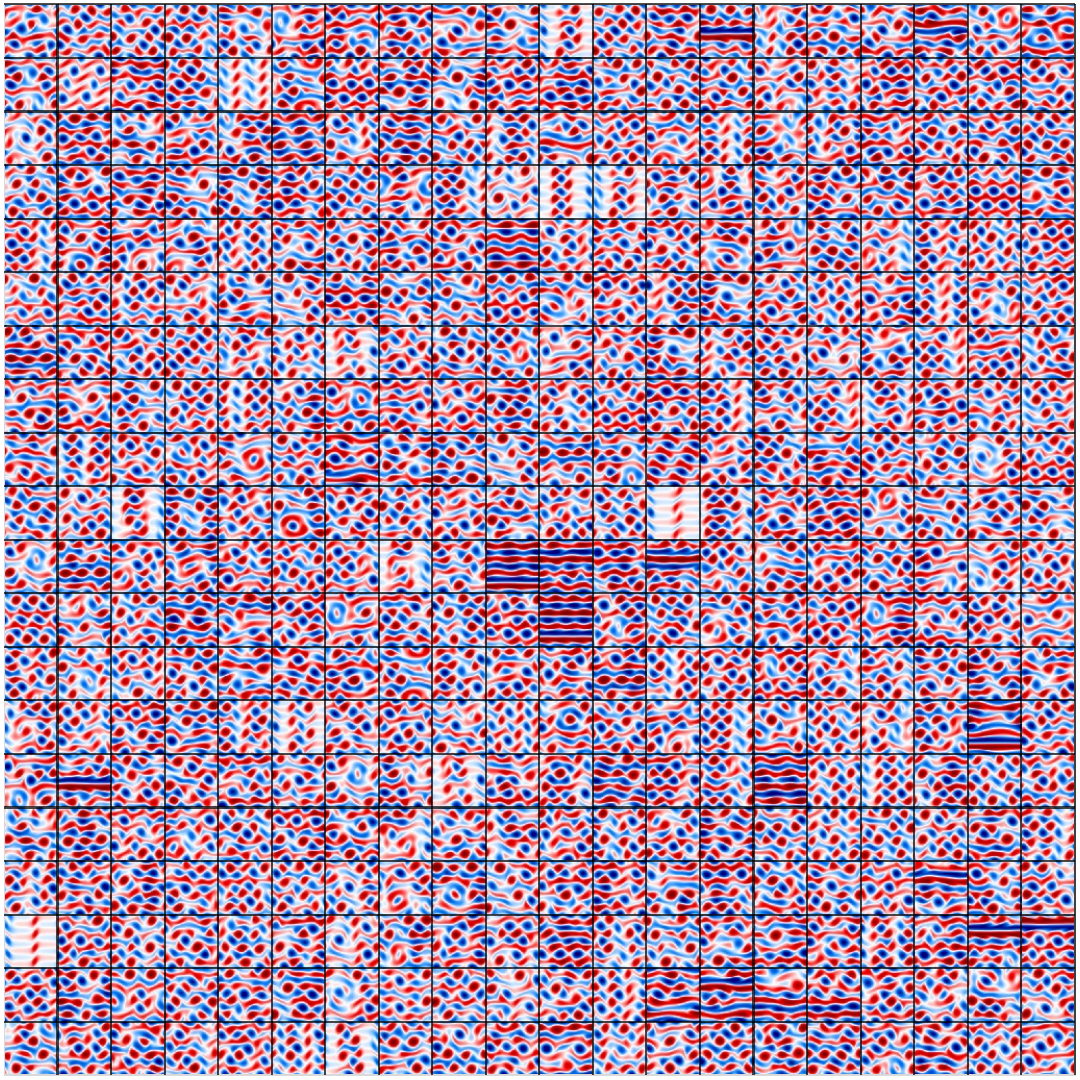}\\
    \caption{A random selection of 400 steady state solutions. Vorticity is shown with color limits $[-10, 10]$. }
    \label{fig:rand400}
\end{figure}

\begin{figure}[th!]
    \centering
    \includegraphics[width=\textwidth]{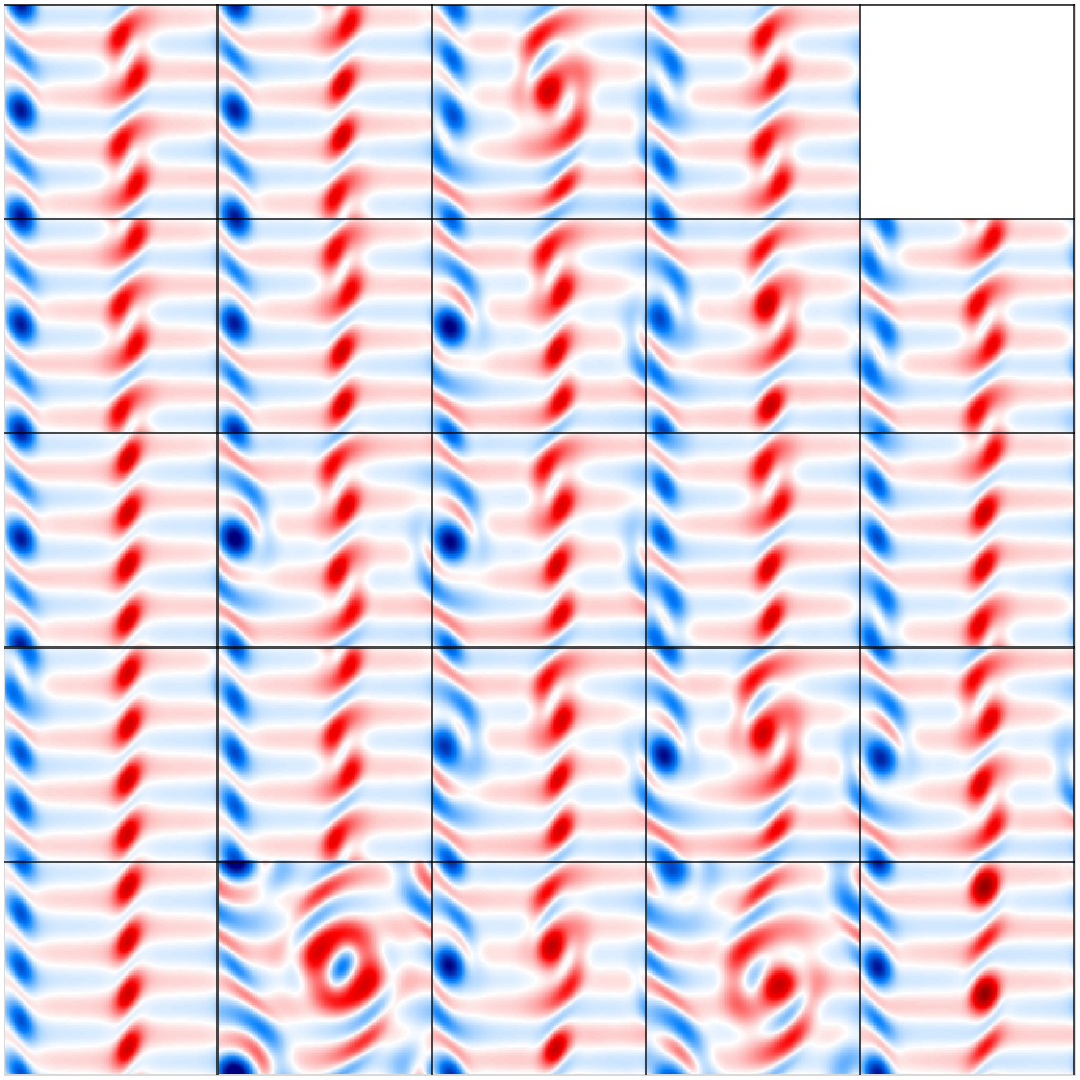}\\
    \caption{The 24 solutions making up Family 1.}
    \label{fig:Fam1}
\end{figure}

\end{appen}

\end{document}